# Preliminary Report of the AMS analysis of tsunami deposits in Tohoku – Japan – 18th to the 21st Century

Wassmer, P. [1,2,3] *, Gomez, C. [3*], Hart, D.E. [3], Hiraishi, T.[4], Azuma, R. [4], Koenig, B. [1], Trautmann , M.[1]

[1]*Faculty of Geography and Planning, University of Strasbourg, Strasbourg, France*

[2]*Laboratory of Physical Geography (LGP), University of Paris 1, Panthéon-Sorbonne, France.*

[3]*Department of Geography, College of Sciences, University of Canterbury, Private Bag 4800, Christchurch 8140, New Zealand.*

[4]*Research Center for Fluvial and Coastal Disasters, DPRI, University of Kyoto, Fushimi-ku, Japan.*

*\* Corresponding authors: patrick.wassmer@unistra.fr & christopher.gomez@canterbury.ac.nz*

**Abstract**

Sedimentary records of tsunamis are a precious tool to assess the occurrence of past events, as attested by an abundant literature, which has seen a particular 'boom' in the aftermath of the 2004 Indian Ocean tsunami and the 2011 Tohoku tsunami. Despite an extensive literature, there is very little to no understanding of the role that the changing coastal environment is playing on the record of a tsunami, and for a given location, it is still unclear whether the largest tsunamis leave the largest amount of deposits. To research this question, the present study took place in Japan, in the Tohoku Region at Agawa-pond, because the pond act as a sediment trap. Using a sediment-slicer, a 1 m thick deposit was retrieved, from which 4 tsunami sequences were identified, including the latest 2011 tsunami. Using a series of sedimentary proxies: the AMS (Anisotropy of Magnetic Susceptibility), grain size analysis, quartz morphoscopy (morphology and surface characteristics) and the analysis of microfossils, disparities between the tsunami deposits were identified and most importantly a clear thinning of the tsunami deposit towards the top. Provided the present evidences, the authors discuss that the upward fining is due to at least two components that are seldom assessed in tsunami research (1) a modification of the depositional environment, with the progressive anthropization of the coast, providing less sediments to remobilize; and (2) a progressive filling of the Agawa pond, which progressively loses its ability to trap tsunami materials.

**1. Introduction**

Less than 7 years after the 2004 Boxing Day tsunami devastated Indian Ocean coasts (Lavigne et al., 2009; Paris et al., 2009), the Great East Japan Earthquake 9.0 Mw megathrust earthquake generated the Tohoku-oki tsunami (UTM-54N E505030 - N4267960) at 14:46 local time on 11 March 2011 (Chester et al., 2013; Hwang, 2014; Koyama and Tsuzuki, 2013; Kyriakopoulos et al., 2013, Mori et al., 2011). The tsunami waves swept along a 2000 km stretch of the east coast of Japan (Ramirez-Herrera and Navarrete-Pacheco, 2013) and impacted shores as far afield as North America (Thomson et al., 2013). For better or worse, this catastrophic event (e.g. Schwantes et al., 2012) has generated an extremely high number of scientific publications (Gomez and Hart, 2013; Gaillard and Gomez, 2014), including numerous contributions on the ensuing erosion and sedimentation of the Sendai Plain. These studies indicate that very few diatoms found in the tsunami deposits belonged to truly marine species (Szczuciński et al., 2012; Takashimizu et al., 2012), and that the mineral composition points to shallower, more coastal source areas (Nakamura et al., 2012), and sometimes precise nearshore sand origins (Jagodzinski et al., 2012). Such results have also been confirmed by ostracod analysis, indicating shore-based sediment origins (Tanaka et al., 2012; Naruse et al., 2012). Significantly, a detailed, localized study of a multi-kilometer transect supports the idea of one long, shore-parallel, coast-proximal source of tsunami deposit sediments; that is, the results evidence the shoreline and terrestrial origins of the material (Putra et al., 2013), as opposed to deeper marine origins. The dominance of locally-sourced sediments in tsunami deposits was also reported on the northern shore of Sumatra after the 2004 Indian Ocean Tsunami (IOT) (Wassmer et al., 2008). Therefore, this suggests that the position and amount of tsunami deposits are not only linked to the energy of the tsunami waves, but also to the type of landcover near shore, and how the tsunami waves have behaved once on land. Using previous tsunami deposits archives, one should therefore observe non-linear relationships between the deposits and the magnitude of the tsunamis: because the coast was less populated with a different landcover; because the coast have had an evolving configuration; and because the waves have travelled following different patterns depending on the source of the tsunamis…

Consequently, the present contribution aims to shed new light on the historic tsunamis of the Tohoku coast via an investigation of sedimentary records of the northern Sendai Plain and it aims to show that the evolution of the coast should – in theory – play an important role in shaping the characteristics of tsunami deposits.

While the 2011 tsunami has been the specific focus of a lot of recent scientific attention – including over 2,000 references in Scopus® – it is well-known that *several* tsunamis have swept the same coast historically (Liu et al., 2012). During the last 120 years, the 1896 Meiji tsunami, the 1933 Showa tsunami, and the 1960 Chilean tsunami pre-date the 2011 disaster, with the 2011 and 1896 events constituting the first and second most destructive events over this period (Liu et al., 2012; Kanamori, 1972).

The impacts of the Meiji tsunami (15 June 1896) were particularly severe on the coast near Rikuzentakata (UTM N54 E554090 – N4319950) in large part due to the arrival of the waves coinciding with the local high tide. This effect was pronounced, despite the area having a microtidal regime with a tidal range of around 1.5 m. The inundating tsunami wave run-up height reached approximately 4.6 m above sea level and inundated an area of 1.56 $km^2$ (Matsuo, 1933). The Showa tsunami (3 March 1933) wave run up was smaller than that of the Meiji tsunami, reaching between 3.5 to 3.8 m above sea level and inundating an area of 1.34 $km^2$ (Matsuo, 1933). The Chilean tsunami of 22 May 1960 was a far-field tsunami, triggered by a 9.5 Mw earthquake in Chile, and inundating more than 5 $km^2$ of the Rikuzentakata coast area, with wave run up of 4.5 to 5 m. Most likely, the extent of the inundation of this event was greatly increased due to the wave period matching the natural resonance of the bay (Satake and Kanamori, 1991). Finally, on 11 March 2011 the Tohoku-oki tsunami inundated an area of about 13.45 $km^2$ and produced wave run up heights 15 m above sea level – 0.2 m at low tide.

The present research was conducted on Honshu Island, Japan, in the Tohoku region (literally meaning the 'east–north' or northeast) from which the 2011 tsunami takes its name. The Sendai coastal plain is located to the east of Eponym City, in the province of Miyagi, north of Fukushima prefecture and south of Iwate prefecture. The area is characterized by a long stretch of sandy shoreline, which ends in the north with a series of pocket beaches embayed by rocky headlands. At the junction between these two beach types lies the coastal village of Shiogama, including Agawa Pond, the feature from which the material used in the present study has been retrieved.

## 2. Materials and methods

### 2.1. Sampling for sediment analysis

The sedimentary series used for the present study has been acquired from a core slice of the bed of Agawa Pond (in eastern Shiogama, Miyagi prefecture – Fig. 1), which was drained in the aftermath of the

2011 tsunami. The sediment core was retrieved using a geo-slicer (Fukken Co. Ltd.), which is a coring technique developed in Japan that creates very little to no disturbance in the grain fabric of fine deposits as they are sampled. The system is made of a rectangular casing that cuts a slice out of the subsurface. Once retrieved, the casing is then opened for sediment sampling and analysis.

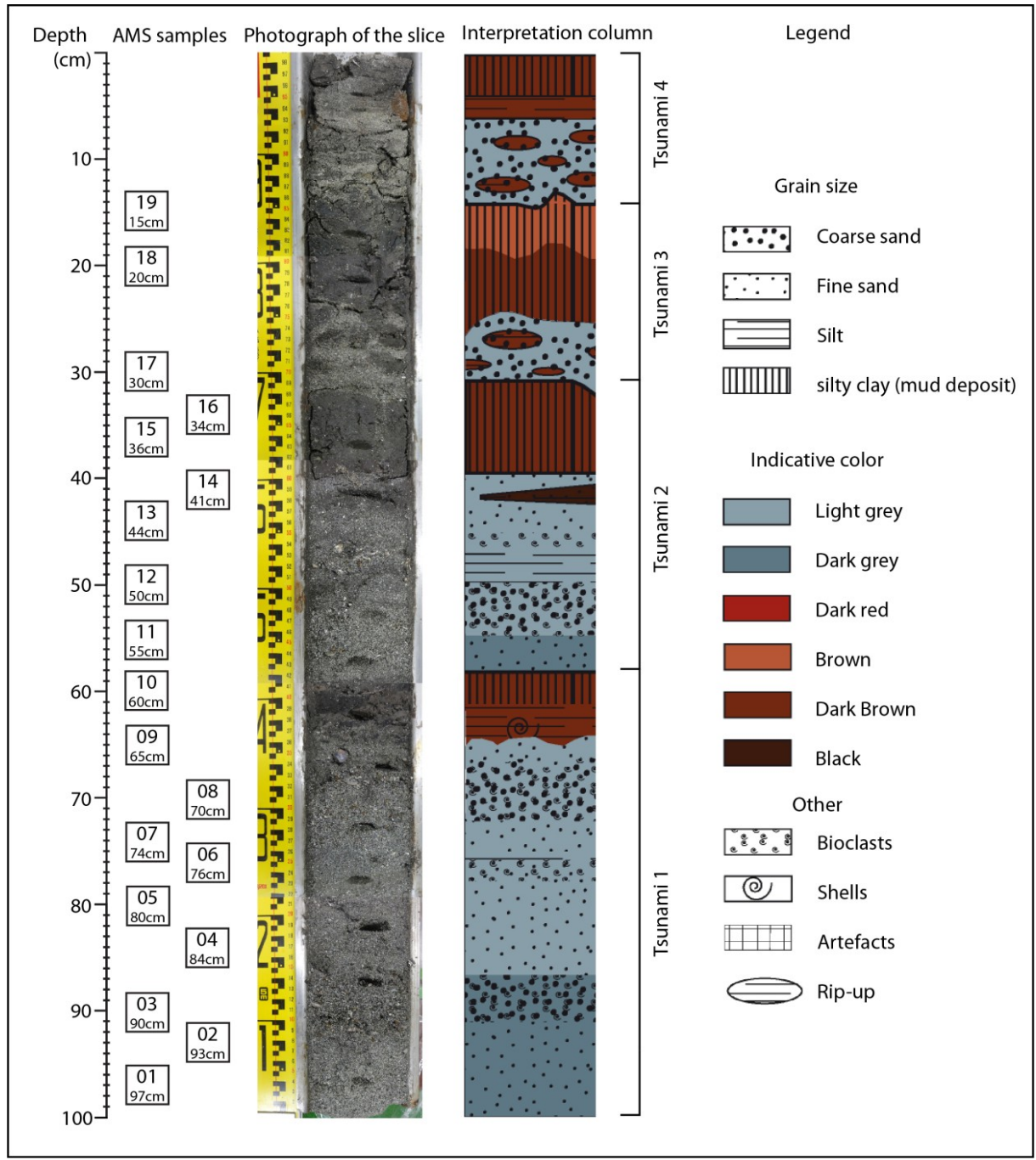

Fig. 1 Core from Agawapond with the samples.

The rectangular core retrieved with the geoslicer was 9.6 cm wide, 3 cm thick, and 100 cm long. Initial visual observations along its length revealed the existence of four similar, superimposed, sequences: these were labeled from the base to the top of the core as Ts1 to Ts4. Each sequence comprised sub-layers of fine to coarse sand containing bioclasts in various proportions and capped by a layer of dark-brown sandy silt. Given their characteristics and similarities, we believe that each sequence in the core corresponds to a single tsunami event. The last sequence at the top of the core, Ts4, most likely corresponds to sediments emplaced by the 2011 Tohoku-oki tsunami.

A total of 39 samples were taken from the core for sediment analyses as follows: (a) systematic sampling of the core every 5 cm produced 20 samples, and (b) a further 19 cube samples were taken for AMS or Anisotropy of Magnetic Susceptibility analysis (technique details below) and these latter samples are denoted throughout this paper by an asterisk.

Three complimentary methods of sediment analysis were applied to the Agawa Pond core: (1) an electromagnetic method developed by the authors for interpreting tsunami deposits - the AMS for soft sediments (Wassmer et al., 2010; Wassmer & Gomez, 2011); (2) foraminiferal data extraction; and (3) grain size and shape characteristics measured using a Shimatsu SALD (Series Laser Diffraction Particle Size Analyzer). In addition, further interpretations were produced using diachronic aerial photograph analyses, specifically structure-from-motion and multiple-view stereo-photogrammetric techniques (Gomez, 2014a,b; Gomez, et al., 2014).

**2.2. The Anisotropy of Magnetic Susceptibility Method**

The anisotropy of magnetic susceptibility (AMS) is an electromagnetic method that uses the time an induced electromagnetic signal takes to travel through sediments to detect the magnetic alignment of deposit crystals. The AMS has been traditionally used on rock dynamics to infer emplacement mechanisms and orientations (Hamilton and Rees, 1971; Ellwood, 1980; Tarling and Hrouda, 1993; Borradaile and Henry, 1997; Bradak, 2009). More recently the method has been adapted for use on unconsolidated, sandy tsunami sediments, in order to retrieve the preferential orientation of the crystals in sand grains and, thus, the sediment fabric (Wassmer et al., 2010; Wassmer & Gomez, 2011). It is possible to use this property as a proxy of the orientations of grains in a deposit. According to Runsack (1957), when clastic sediments are emplaced, the long axes of the sand grains tend to align themselves parallel to the fluid flow direction. Runsack (1957) explained that the more elongated sand grains are, the more parallel to the fluid-flow direction they tend to be. Also, the grains tend to be imbricated (shingled) according to the direction of dip upstream. Additionally, he specified that the fluid-flow direction of the depositing medium can be determined from an orientation analysis of the dimension fabric of a sand deposit. Evidencing sedimentary fabric and, thus, the flow direction is the major contribution of the AMS technique. Depending on the shape of the grains, their concentration in the traction carpet and the flow velocity, elongated particles can be imbricated (their long axis showing the flow direction) or can roll on the bottom (their long axis aligned perpendicular to the flow direction). To address this complexity, and avoid potential misinterpretations in inferring the flow direction from the dip and azimuth of the long axis of the ellipsoid of anisotropy (*Kmax* or $K_1$), the present contribution uses the "*Kmin* or $K_3$", which is the pole of the magnetic foliation following Palmer et al. (1996), Ort et al. (2003), Giordano et al. (2008), La Berge et al. (2009) and Wilcock et al. (2014).

Parametric values extracted from this method are commonly: the magnetic foliation (*F*), the magnetic lineation (*L*), the shape parameter (*T*), the corrected degree of anisotropy (*Pj*), the alignment parameter (*Fs*), and the ellipsoid shape factor (*q*)

Parameter *F* (foliation) reflects the intensity of the planar linear orientation, while *L* (lineation) characterizes the intensity of the planar parallel orientation. A high *F* value corresponds to a deposit transported as traction load (for an environment where the fluid isn't still and characterized by a current) while a high *L* value reflects a deposit process dominated by settling. The alignment parameter *Fs* increases with the energy of bottom currents (de Menorcal 1986, Park et al, 2000). *Fs* values have been used in conjunction with particle size analysis as an indicator of palaeo-current velocity variations by Elwood and Ledbetter (1977). The shape parameter *T* distinguishes the relative dominance of traction versus settling as influences on the resulting deposit: when *T* values are between 0 and 1, settling dominates (1 indicates a purely foliated deposit) whereas when *T* values are between 0 and -1, lineation dominates (-1 indicates a purely lineated deposit). *Pj* is typically a measure of the eccentricity of the anisotropy ellipsoid. If *Pj* = 1, the fabric is perfectly isotropic; between 1 and 1.05 the fabric is weak while between 1.1 and 1.2 it is strong. The ellipsoid shape factor *q* is an indicator of either depositional or tectonic fabric. A value less than 0.7 is indicative of an undeformed sediment fabric (primary fabric) while a value higher than 0.7 suggests that a tectonic activity deformed the primary fabric (Hamilton and Rees, 1970).

An AMS core sampling strategy was established to plug the sampling boxes into the center of units and avoid unit ends and transitions, in order to avoid the pollution of one deposition process over another linked to shear stress and material mixing. Nineteen AMS samples were analysed using the AGICO MFK1A Kappabridge® with automatic spinner, following the same method adopted in Wassmer et al. (2010); Wassmer and Gomez (2011) and Kain et al. (2014). Recently, the AMS has also been referred to as 'the Magnetic Fabric' MF (Kain et al., 2013), but the present contribution keeps the first terminology widely used in the scientific community: AMS (Wassmer et al., 2010; Wassmer & Gomez, 2011).

N.B.: Samples collected too close to the interface between tsunamis 2 and 3 (28* & 29*) as well as between tsunamis 3 and 4 (36*) were excluded from this analysis due to the potential mixing of fine material from the top of the older (lower) tsunami sequence with coarse material from the base of the younger (upper) tsunami sequence in these samples.

### 2.3. Grain-size, quartz morphoscopy and foraminifera analyses

To complement the AMS results, which offers data on wave orientations, the authors also analyzed the sediment units for grain sizes, quartz morphoscopy (morphology and surface state) and foraminiferal species in order to ascertain the environmental origin of the sediments. From the grain size analyses, parameters were extracted: mean grain size according to Trask (µm); standard deviation (Φ) calculated according to the methods of Folk and Ward (1957) and Blott and Pye (2001); skewness according to Trask(µm). To confirm the observations and the general data obtained by the described methods, we used Passega's CM diagram according to the technique of Passega and Byramjee (1969), and Passega

(1977). The mean grain size was plotted on a log/log diagram with the fifth percentile C95 (µm). This provided further insights into the deposition conditions experienced by the different tsunami units

Sand samples were washed under distilled water on a 50µ sieve and dried in an air oven. Each sample was homogenized and submitted to a double quarting. From the remaining sands, a random sample of 100 grains was studied under a binocular microscope using the quartz morphoscopy method, which focus on the grain shape (angular to well rounded) according to Pettijohn (1957) and on the grain surface characteristics (polish / matt, presence of coating) according to Tricart and Cailleux (1963), in order to identify the main processes responsible for the sediment transport.

Foraminiferal analysis is a precious tool in geology and stratigraphy. With contemporary deposit investigations, such as those examining tsunami, information about the composition of foraminifera assemblages may be analysed from to differentiate sediments sourced from marine versus terrestrial versus estuarine origins and even to detect the water depth from which the sediments were entrained or the distance of transport before deposition (Chagué-Goff et al., 2011). Changes in assemblage composition within a sedimentary sequence indicate changes in marine environmental conditions. Species with restricted environmental niches are particularly instrumental in such palaeogeographic analyses and palaeoenvironmental reconstructions (Chagué-Goff et al., 2011). For the present study, 569 specimens were examined and the dataset was studied in light of previous work by Griveaud (2007), Hussein et al. (2006), Nanayama and Shigeno (2006), and Uchida et al. (2010). To classify the origin of the foraminifera, the authors have adopted a depth-based division in three 'zones', following Nanayama an Shigeno (2006): i.e. an inner sub-littoral zone (ISZ) with water depths up to 50 m, a medium sub-littoral zone (MSZ) with water depths between 50 and 90 m, and an outer sub-littoral zone (OSZ) with water depths between 90 and 240 m. The foraminiferal investigations were carried out based on 100 g of sediments taken from homogenized samples placed in an experiment tray. The foraminifera were removed one by one from the rest of the sediments using fine tweezers and a binocular microscope, in order to conduct the species identification. This operation was repeated three times for each sample.

Examples of the species found were then photographed using the microscope software *Motic Images Plus 2.0 ML*.

## 3. Results

The in-situ observations of the extracted core revealed a series of four tsunami deposits, corresponding to the four last tsunamis that inundated the area during the last ~150 years. The first sequence, Ts1, located at the base of the core between 100 and 58 cm below the core surface, was characterized by a series of sandy layers with bioclasts topped with a sandy-silt unit (Table 1).

*Table 1. Detailed description of the sediment deposit sequences observed in the Agawa Pond core*

| Deposit sequence | Depth below core surface (cm) | Layer description |
|---|---|---|
| Ts1 | 100-91 | well sorted grey sands |
|  | 91-86 | lightly erosive contact at base, then coarse grey sands containing abundant bioclasts, oriented into roughly horizontal layers |
|  | 86-74 | Relatively homogenous finer sands with small bioclasts, no visible layering except a thin, grey, very fine sand layer (78-77 cm), resting atop a line of abundant bioclasts. |
|  | 74-67 | coarse sands with omnipresent bioclasts, overlain by unbroken shells followed by an undulating contact underlined |
|  | 67-64 | greyish fine sands, evolving into next layer |
|  | 64-58 | dark grey silts poor in bioclasts including elongate clasts of darker, fine material |
| Ts2 | 58-59 | erosive contact |
|  | 59-52 | medium to coarse sands, lighter grey in color and containing numerous bioclasts, evolving into next layer |
|  | 52-48 | finer and darker sands |
|  | 52/48-45 | coarse sands with abundant coarse bioclasts, strongly erosive of deeper sub-layer which was locally erased by an undulating contact |
|  | 45-40 | elongate clasts of brownish silty material |
|  | 40 to 32 | dark brown silts characterized by a lack of bioclasts |
| Ts3 | 32-25/24 | clean, grey coarse sands with bioclasts present but not abundant, inclusions of large (3-4 cm length) dark-brown silt clasts |
|  | 25/24-14 | dark-brown compacted silt and clay, devoid of bioclasts |
| Ts4 | 14/13-10 | erosive contact with the previous sequence, followed by clean, grey, medium sized sands containing numerous dark-brown silt elongated rip-up |
|  | 10-8 | dark silt interleaved by thin, clean, sandy beds |
|  | 8-4 | Well sorted layer of clean, grey sands with little bioclast content |
|  | 4-0 | a large rust-colored artifact (~3.5 cm diameter) separated, flowed by dark-brown clayey silts |

For this earliest deposit the information collected was partial since its base was too deep to be sampled with the 1 m geo-slicer. Despite the absence of its base, this tsunami deposit was still significantly thicker than the three more recent ones above.

The second tsunami deposit, Ts2, was found between 59 and 32 cm below the core surface (Table x). The third tsunami deposit, Ts3, was situated between 32 cm and 14 cm below the core surface while the fourth (Tohoku-oki) tsunami deposit, Ts4, lay between 14 and 0 cm below the core surface. Sandy layers topped by sandy-silt deposits characterized all four tsunami deposits sequences sampled. The presence of numerous coarse dark-brown rip up-clasts in the sandy sequences were a characteristic specific to the last two tsunami deposits, Ts3 and Ts4. All four observed sequences were characterized by a progressive decrease in thickness towards the surface, with total sequence thicknesses equaling 42, 26, 18 and 13-14 cm from the core base to top.

### 3.1 Grain-size and grain morphoscopy

The grain-size distribution was established for each sample layer within the successive sequences. Within each of the four sequences Ts1 to Ts4, the base comprised coarse sands (500 to 700 µm), with the texture fining upward into a layer of fine sands (260 to 240 µm). The proportion of sand in this part of the deposit varied from 100 to 97%. A very small amount of silt (2-3%) was present only at the very end of these sand-dominated layers. The top of each tsunami deposit sequence comprised a layer of dark-brown material (221 to 64 µm) where silt dominated and the sand content decreased to values comprised between 30 and 17%, with less than 10% clays.

Within Ts1, three superimposed upward fining sub-sequences were discernable, albeit with the first sub-sequence having an incomplete base. Within Ts2, two normal graded sub-sequences were superimposed. Ts3 and Ts4 were each characterized by a single fining upward layer. In addition to the within-sequence textural changes, when comparing textures between sequences there was a clear trend of decreasing mean grain sizes from Ts1 to Ts4.

The evolution of grain size sorting within the core revealed similar patterns in all four tsunami deposit sequences. The general tendency was for good sorting at the base of each sequence, with (values close to 0.5), followed by a reduction in sorting, with moderate to poor sorting (values >2) within the fine unit at the top of each sequence (Fig. 2). These patterns are consistent with the results obtained by Szczuciński et al. (2012) for the Tohoku-oki tsunami deposit on Sendai Plain. Within the two first sequences, the two steps evolution evidenced in the grain size are reflected in the sorting values, which ranged between 0.9 – 1.0 at the base and 1.6 – 1.7 at the top. The skewness values are all above 0, indicating that all the sequences were skewed in favour of coarse particles (Fig. 2).

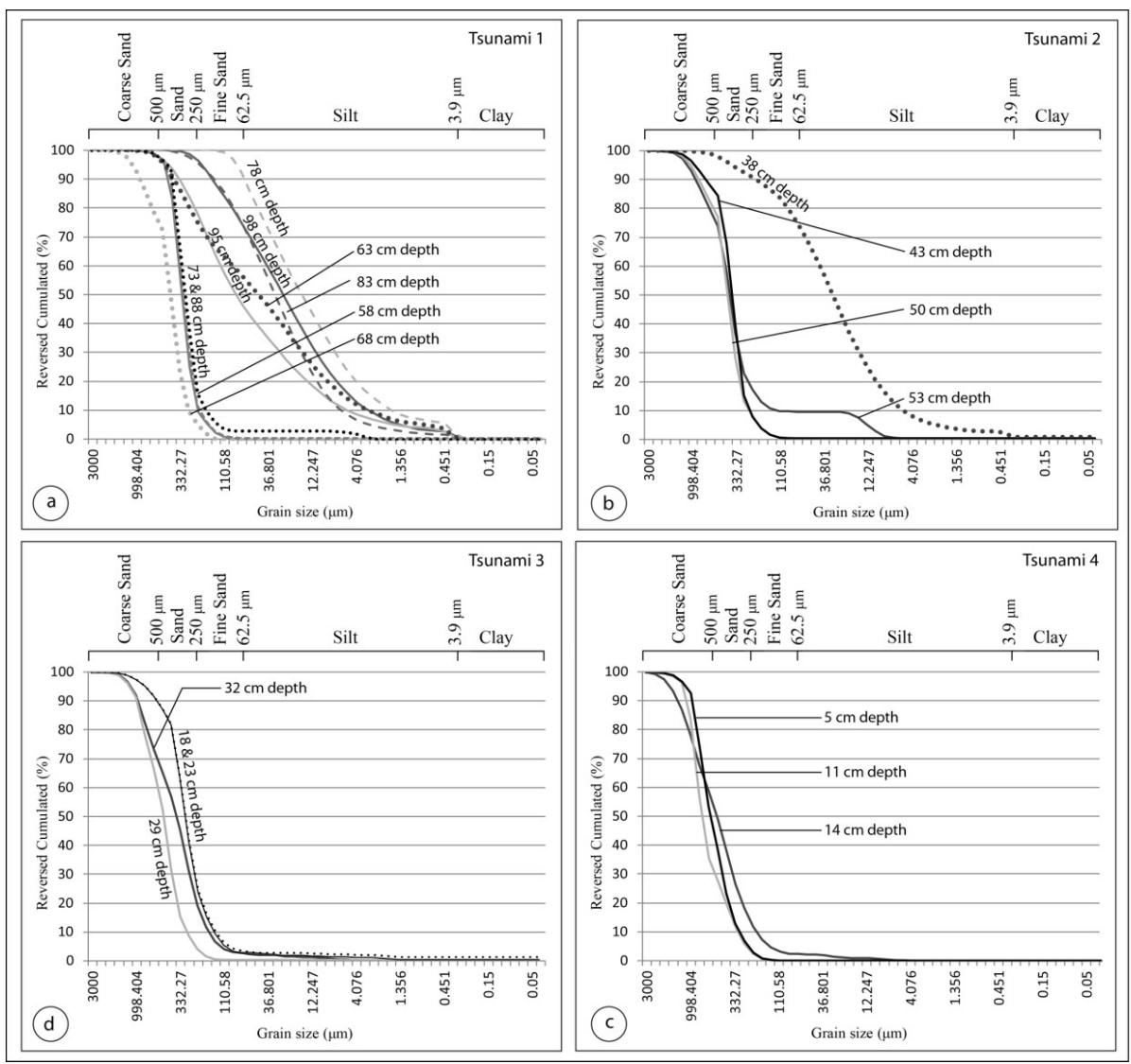

Fig. 2 Grain-size distribution

The morphoscopy of the sands shows a clear predominance of highly polished grains in the material emplaced in Agawa Pond. The proportion ranged from 70 to 89% within the first sequence, to 57% in the fine material constituting the end of the sequence. This polishing indicates a littoral origin (s.s), nearshore or beach, for this material polished by the back and forth action of waves. A smaller proportion of the sandy sediments (15-20% at the base of the core, 25-35% at the top) exhibit surfaces pitted with thousands of microscopic cavities. These visually matt grains evidence sediments originating from local aeolian onshore deposits such as in the beach backshore and dunes. Generally the grains in the core were found to be well rounded to rounded (50-70%). The proportion of blunt grains was generally around 20 to 30%, but increased in the muddy layers of each sequence end, reaching 80% at the top of the first sequence. These grains were typically yellow or rust colored.

The proportion of bioclasts and microfossils varied throughout the whole core. Their concentrations changed simultaneously, decreasing or being absent in the fine material marking the end of each sequence.

**3.2 Analysis of deposition characteristics from the Passega CM diagrams**

Based on the analyses using CM diagrams , the four investigated tsunami sequences exhibited a range of sediment deposition characteristics that varied from rolling to uniform suspension, with most of the material having been deposited via rolling and gradual suspension. Keeping in mind that we did not recover the full first tsunami deposit, the sediments of Ts1 (Fig. 3-1) displayed a series of alternating deposition modes, varying between rolling and gradual suspension from the base of the core upwards, before finishing at the top of the sequence with sediments reflecting uniform suspension characteristics. The mean grain size was coarse (200 to 725 µm), except for the uppermost sequence sample, which had a mean grain-size of 65 µm. Ts2 (Fig. 3-2) exhibited a similar median grain size distribution, with sample mean grain diameters ~ 522 µm, except for the very top of the sequence, which had a mean grain size between 40 and 50 µm. As in Ts1, Ts2 displayed alternating layers with rolling and ground suspension characteristics, topped by a layer reflecting uniform suspension. The characteristics of Ts3 and Ts4 were simpler, these deposits being thinner than the previous ones and showing less layering (Fig. 3-3,4), but they too displayed the same general pattern of coarse material at the bottom of the column and a layer of fine material at the top of each sequence.

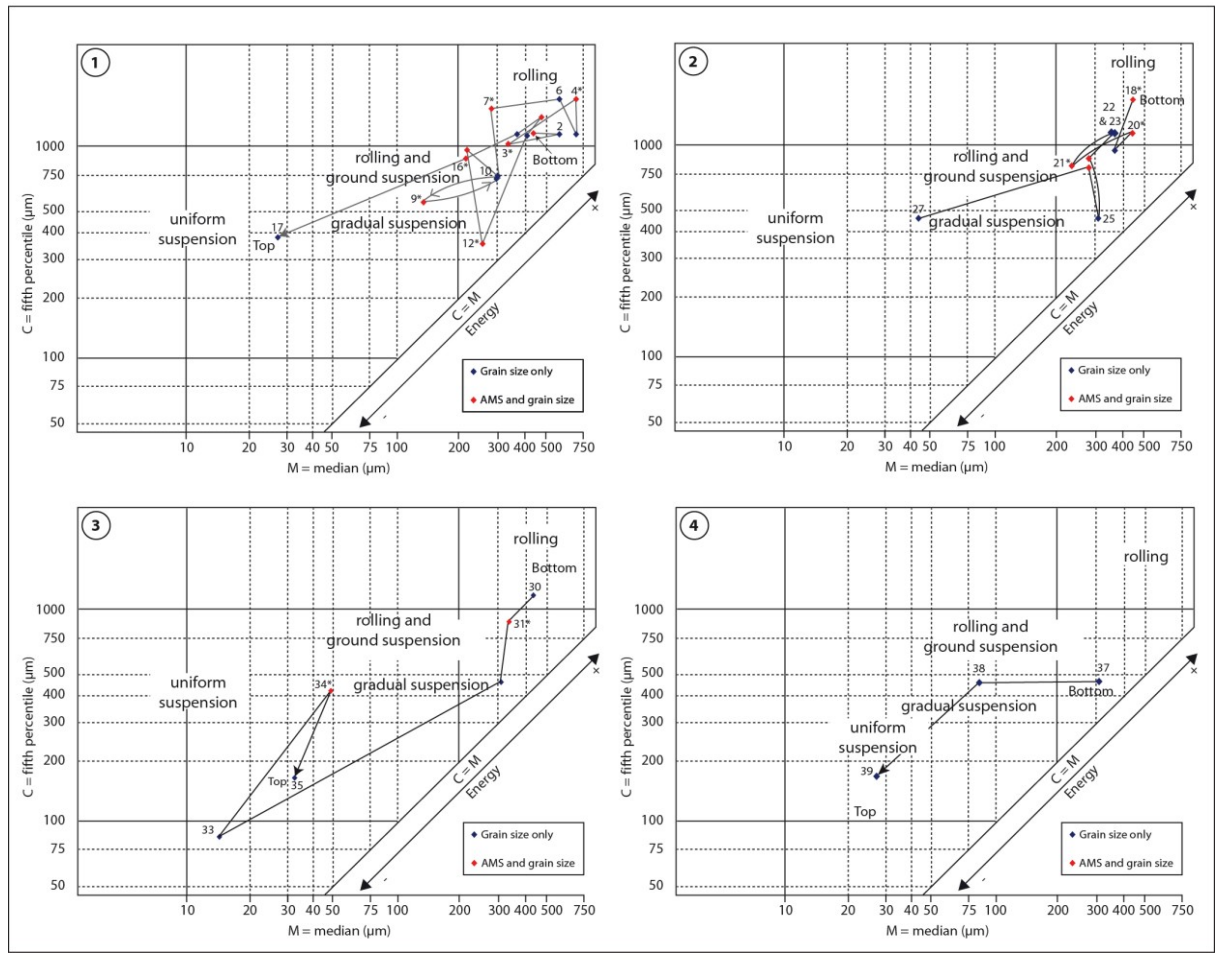

Fig. 3 Passega Diagram

## 3.3. Anisotropy of Magnetic Susceptibility parameters and inferred flow directions

As outlined in the methods section, parametric values are derived from the three axes $(K_1) > (K_2) > (K_3)$ of the anisotropy ellipsoid and these values, combined with the grain size data and the flow direction information, allow us to interpret the hydrodynamic conditions that existed during sediment deposition. For the complete Agawa Pond core, the $Pj$ values indicated a consistently weak to moderate eccentricity of the anisotropy ellipsoid with values ranging from 1.017 to 1.112. The $q$ values (ellipsoid shape factor) were comprised between 0.06 and 1.039 with 10% of the values exceeding 0.7, indicating that the general magnetic fabric acquired during deposition (primary fabric) has not been disturbed by tectonic, bioturbation or compaction (Rees 1961). The first tsunami sequence was characterized by a strong *Fs* value at its base, with a progressive weakening of *Fs* values above. A small peak in *Fs*, with a value around 74, occurred at the base of an uprush phase, and there was a weak increase the *Fs* values at the very end (top) of Ts1.

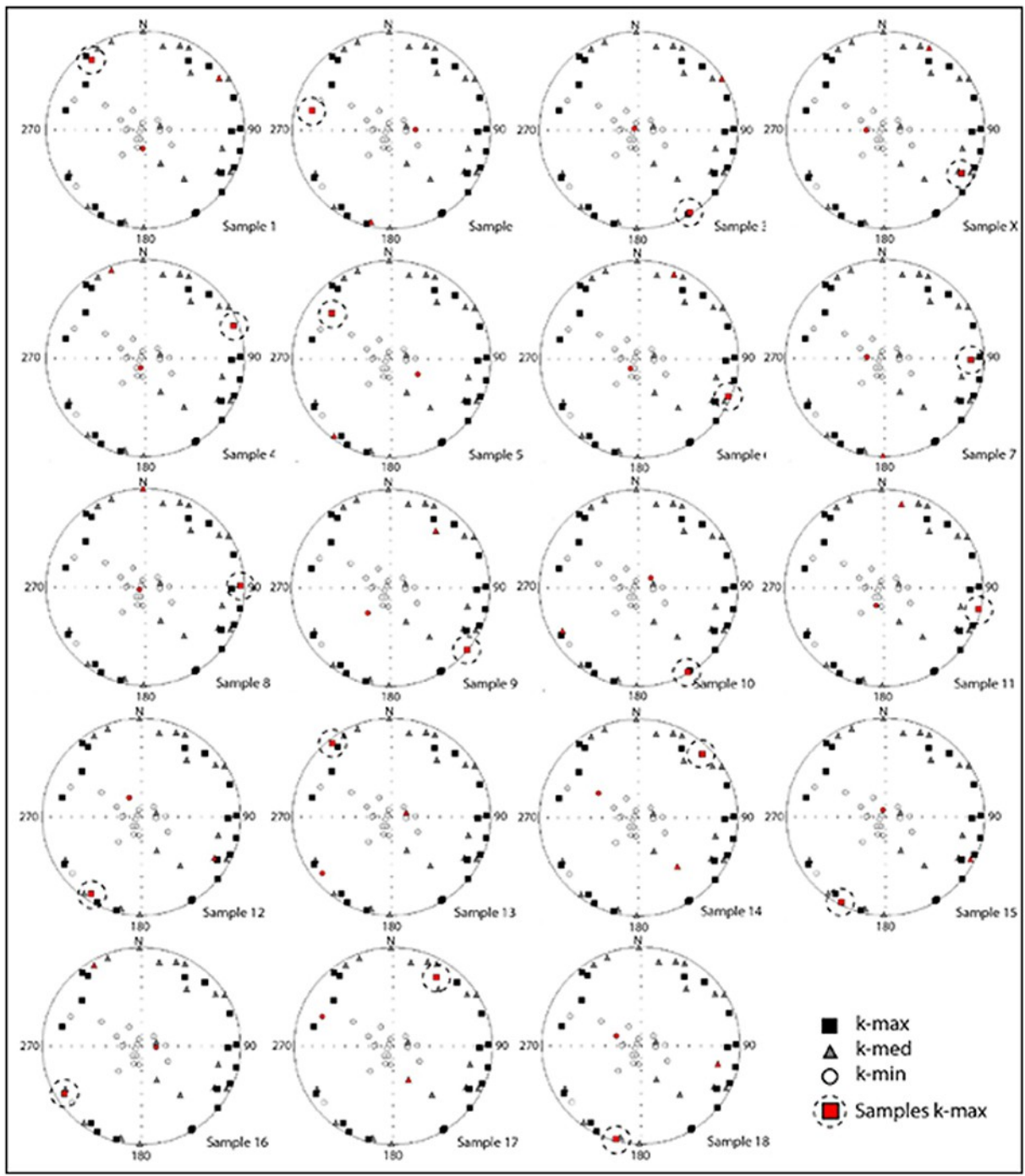

Fig 4 Orientation of the k-max, k-med and k-min for each sample.

. In Ts2 and Ts3, the general pattern of decreasing *Fs* values from bottom to top was found as well, but interrupted by the presence of small peaks of stronger values reflecting punctual short energy recovery. According to the shape parameter (*T*), the three tsunami sequences were largely dominated by settling processes, with the only exceptions being the traction-dominated coarse sands at the base of the core as well at the end of the second sequence.

Topographically, Agawa Pond is a depression starting 350 m from the ocean shore. It is 110 m in width and stretches 700 m inland along a N129º to N309º (SE-NW) axis. This narrow gutter acts as a tsunami sediment trap. It channels tsunami wave uprush, which surges inland in this area at N303º, on a similar angle to the pond long axis. The channelling influence of the pond removes from its deposits much of the sediment emplacement flow direction variations that are observed elsewhere across the surface of the plain. Some tsunami inflows related to the local hilly topography around the pond generate bottom currents with systematic preferential directions (toward the SW for instance). On the one hand, when reconstructing the characteristics of the flows that have emplaced the pond deposits, care must be taken to interpret these findings within the context of the pond's effect on tsunami wave flow directions and not simply as indicative of flow patterns on the surrounding coastal plain. On the other hand, the number of tsunamis recorded in the pond, the characteristics of the sedimentary sequences themselves, and the characteristics of the source material are not simply 'pond-dependant' but rather have implications for the tsunami event effects on the wider coastal plain.

The first tsunami sequence recorded in our core, Ts1, was the most complex. As mentioned earlier, the base of our core was not the very base of the tsunami deposit. In the core, Ts1 begins with sediments indicating a flow oriented towards the SW (100-91 cm from the top of the core). This is consistent with backwash coming off the slopes of the hills located NE of the pond. This backwash emplaced deposit is followed by a coarse deposit (from 91-86 cm) indicating flows oriented towards the SE, consistent with flow channelization along the long axis of the pond towards the coast. The next sediment layer (from 86 to 74 cm) indicates an uprush phase oriented NE, consistent with flows guided inland along the pond gutter. This is followed by another two backwash indicative layers: the first oriented towards the SW

(from 86-78 cm), and the second, guided by the pond, oriented towards the SE (from 78-74 cm). The next deposit layer (from 74 to 67 cm) indicates emplacement by a new uprush phase, with grains oriented towards the NW. Like before, this uprush deposit is followed by two backwash deposits, the first indicating flows oriented towards the E and the second indicating SW-oriented flow corresponding to the deposition of the dark-brown fine materials, marking the end of the Ts1 sequence.

Ts2 begins with an uprush deposit (from 59-52 cm), emplaced by flows heading towards the NE, followed by a layer (from 52-45 cm) emplaced by backwash flows directed towards the SW. The second uprush layer (from 45-43 cm) indicates flows oriented towards the NW and is followed by a backwash layer (from 43-40 cm) indicating SW-moving flows. The sequence ends with a thick muddy layer (from 40-37 cm), deposited by a flow towards the NW.

The Ts3 deposit began with a sediment layer emplaced by an uprush phase that flowed towards the NE (from 32-24 cm) and ended with dark silts (from 24-14 cm) deposited by an uprush flow oriented towards the NW.

**4. Evidence synthesis**

**4.1 Deciphering the hydrodynamic conditions of the four recorded tsunami events**

Agawa Pond has existed for well over a century. During this time, the pond has functioned to record and preserve the sedimentary signatures of four large tsunamis that have surged across the Sendai coast and plain. The multi-proxy approach to describing the pond deposits detailed in this paper now enables us to reconstruct the characteristics of these events and to compare them.

As mentioned earlier, the very base of Ts1 was not captured in our core. This first tsunami event, as recorded in the base of the Agawa pond core, appears to have been very strong considering the dominance of traction processes identified in sediments here and given the large mean grain size (200 to 725 µm) of the materials emplaced. Three fining-upward sub-sequences were identified within Ts1 via textural analysis but, interestingly, the AMS analysis revealed that these were not emplaced by flows coming from a uniform, seaward direction. AMS results revealed that the first Ts1 normal –graded sub-sequence in our core was emplaced by an initial flow heading SW (seaward), followed by another flow heading SE (seaward in alignment with the long axis of the pond). These directions are consistent with backwash flows. Atop these two coarse sand layers, there was a second, apparently fining-upward sub-sequence comprising coarse sands overlain by fine sands. AMS analysis of this sequence showed, however, that the base of this subsequence was emplaced by a seaward-moving flow while the finer material above was emplaced by a land-ward moving uprush flow aligned NW along the gutter-like long axis of the pond. As such, AMS reveals that this second 'sub-sequence' was not in fact a single graded sub-sequence but rather two distinct layers deposited by waves travelling in opposing cross-shore directions. Another sub-sequence comprising an uprush (inland-moving) layer was then identified at the top of Ts1, indicating emplacement by a flow heading in a NW direction. This was followed by deposits indicating a weak current travelling westward (quasi shore-parallel), then towards the south-east (seaward), these 'backwashes' resulting in the deposition of the fine dark brown silts that cap Ts1. For Ts1, the Passega's CM diagrams clearly indicate the dominance of high energies and rolling processes at the base of each of the three sub-sequences, transitioning towards the top into a deposit indicative of lower energies dominated by 'rolling and ground suspension' to 'uniform suspension' at the very top of Ts1.

Ts2 begins with medium to coarse sands deposited on a landward-moving uprush, heading NW. This uprush deposit was followed by a layer of coarse sands, containing numerous coarse bioclasts, that was likely deposited by a quick, strong and energetic backwash heading towards the SW. A second uprush-deposited sub-sequence, slightly discordant with the layer below, was likely deposited by a flow heading towards the NW. The latter uprush was subsequently interrupted by a backwash flow, heading towards the SW, which deposited medium sands. Capping these Ts2 layers, dark muds were likely emplaced by a lower-energy landward-moving flow, the velocity of which progressively decreased as attested by the magnetic foliation pole ($K_3$) coming close to the vertical. This sequence end corresponds to a slackening of wave energy. The two Ts2 sub-sequence 'pulses' described are also clearly evident on the CM diagram, which shows the processes dominating the first pulse evolved from 'rolling' to 'rolling and ground suspension' while those at the end of the second pulse evolved to 'gradual suspension'.

Ts3 was the thinnest tsunami sequence. It started with sands and silts deposited by a flow that was likely travelling towards the NE. The azimuth data for this layer must be interpreted with caution because the sampling was made difficult by the close matrix of sands and silts. The uprush, most likely guided by the gutter shape of the pond, emplaced a layer of sands, indicating high energies, but this material appears to have been mixed in with, and capped by, a 10 cm thick layer of coarse, dark, rip-up of muddy-silt texture. The latter, 10 cm layer is by far the thickest single layer in the core and was emplaced by a relatively slow flow heading landward (NW). The Passega's CM diagram for this sequence indicates that it likely corresponds to a single pulse, characterised by a clear decrease in energy levels from the base of Ts3, where 'rolling' type processes dominated, towards the top of Ts3, where 'uniform suspension' dominated. The very top of Ts3 is characterized by indications of a short recovery of energy levels (point 17), dominated by 'gradual suspension' processes, followed by a penultimate drop in energy levels.

Ts4, the Tohoku-oki tsunami sequence was not investigated using AMS techniques since the upper part of the core became altered during transport to the laboratory. Ts4 was composed of a 5 cm thick sandy layer, containing abundant rip-up clasts of dark mud. The rip-up clasts were 1 cm thick and often elongated with an undulating disposition. The top of this layer was covered by a continuous 1 cm deposit of dark silts. Above, a thin layer of fine grey sands was capped by another 6 cm of dark silts. The CM diagram shows a three-stage evolution of dominant processes within Ts4: from 'rolling and ground suspension' at the base to 'gradual suspension and rolling' towards mid-sequence, finally to 'gradual suspension' at the top of Ts4.

**4.2 Deciphering the historical tsunami events that emplaced the core sequences**

Using existing research conducted along and near the Sendai coast, and the results of core analyses from this paper, we can try to attribute the observed sedimentary sequences to historical tsunami events. While deposition of the last sequence, Ts4, is indisputably related to the Tohoku-oki tsunami event, determining the events which deposited the three lower (earlier) tsunami sequences in Agawa Pond is slightly more complicated. While numerous papers have been written about the Sendai and nearby coasts, most proved unhelpful in decoding our tsunami event records. Goto et al. (2012), for example, suggest that no historical records exist of large tsunami events impacting the Sendai and adjacent plains over the last thousand years, with the exception of one possible event (the 1611 Keicho tsunami). This finding contrasts with well-known seismically-active nature of the region and the records of several smaller tsunamis having occurred in the area (http://www.bo-sai.co.jp/chirijisintunami.html, http://showa.mainichi.jp/news/1960/05/post-9900.html). The magnitudes of some historical earthquakes known to have occurred in this region, and the likely sizes of associated tsunamis, have been estimated based on the known geological record augmented by numerical modeling studies (Satake et al., 2008; Namegaya et al., 2010; Sugawara et al., 2012, 2011). In their work focused on Rikuzentakata in Iwate Prefecture, Liu et al. (2013) compared the characteristics of the Tohoku-oki tsunami with those of the three previous historical tsunamis known to have impacted the wider east coast of Japan: the Meiji tsunami in 1896, the Showa tsunami in 1933 and the Chilean tsunami in 1960.

Even if the data are scarce and sometimes difficult to compare, a rough classification of historical tsunami events as evidenced by existing literature can be attempted. The Meiji tsunami occurred on June 1896 and was triggered by an earthquake of Mw 8.2 to 8.5. This event coincided with a high tide and was dominated by strong inland-flowing uprush. According to Liu et al. (2013), the Meiji and Tohoku-oki tsunamis can be considered roughly equivalent in intensity. This similarity includes flow depths and run-up levels measured along the coast, with the more recent tsunami representing a slightly larger event: run-ups of 40 m for Tohoku-oki versus 20 to 35 m for Meiji (see Unohana and Oota's 1988 compilation of observations from Iki, 1896; and Yamana, 1896). The Showa tsunami occurred in March 1933, and was generated by a Mw 8.4 earthquake. Run-up levels measured after the event reached 20.3 m in elevation above sea level according to Otuka (1934). With its bore-like waveform, the highly turbulent wave front from this event is reported to have caused severe coastal infrastructure damage (CFICT, 1961). In May 1960 a Mw 9.0 earthquake off the coast of South America generated the so-called 'Chile tsunami', which travelled for 22h and 17 000 km across the Pacific Ocean to the Japanese coast. Due to its extremely long wavelengths, the water flooding the area of Sendai during this tsunami appeared more like a rapidly-cycling, high tide than a series of sea-surface waves. The area of Rikuzentakata, on the Sanriku coast, flooded during this event was more extensive and deeper (5.25 km$^2$/ 3 km) than that of the Meiji (1.56 km²) or Showa (1.34 km² / 1.3 km) tsunamis. With its recorded flood extent of 13.45 km² and upriver maximum inundation of 8.1 km, the 2011 Tohoku-oki event is easily the largest tsunami to have occurred on the east coast of Japan over the 120 last years (Table 2).

Table 2, Overview of likely origins of the tsunami deposits in the Agawa Pond core.

| Tsunami event | TOHOKU-OKI | CHILEAN | SHOWA | MEIJI |
|---|---|---|---|---|
| Date | 11 March 2011 | 22 May 1960 (far-field tsunami reached Japan's coasts after traversing 17 000 km of Pacific Ocean) | 3 March 1933 | 15 June 1896 |
| Earthquake magnitude | 9 | 9,5 | 8,1 | 7,2 |
| Deposit thickness in Agawa Pond (Wassmer et al., this study) | 14 cm | 17.5 cm | 27 cm | >41.5 cm |
| Inundation area at Rikuzentakata (Liu et al., 2013) | 13.45 km² | 5.25 km² | 1.34 km² | 1.56 km² |
| Up-river maximum inundation (Liu et al., 2013) | 8.1 km | 3 km | 1.3 km | |
| Flow height at Rikuzentakata (bay entrance) | ~ 15 m | 2-3 m | 11.8 m | ~ 4.6 m |
| Flow height at Rikuzentakata (bay head) | | 4-5 m | 3.5 m | |
| Wave period | extreme peak: 8 mn superimposed to 30 mn elevated water (Liu et al., 2013) | 60-80 mn (CFICT, 1961) | 10 mn (Matsuo, 1933) | |
| Characteristics | | | Highly turbulent front (CFICT, 1961) | Coincided with high tide level |
| | | | | |

In the Agawa Pond core, we believe that the basal sequence is likely related to Meiji tsunami. The AD869 Jōgan tsunami might have been another possible origin for the sequence except that the volcaniclastic ash layer, called the "AD915 Towada-a tephra", which generally overlies this tsunami deposit elsewhere, making it relatively easy to identify (Goto et al., 2012), did not occur in our Agawa pond core. Given that the pond feature has been shown to be highly effective at preserving deposits, we judge it unlikely that the tephra layer was deposited within pond's sediments and subsequently eroded. If our interpretation of the origin of Ts1 is correct, then Ts2 may be attributed to the Showa tsunami and Ts3 likely corresponds to a Chilean tsunami origin

Although less documented by the international scientific community, the 1960 Chilean tsunami invaded the Sendai plain as well. Houses were destroyed and flooded like the historical photographs of Oofunato City in the province of Iwate (http://www.bo-sai.co.jp/chirijisintunami.html). In the province of Miyagi, there are report of 312 houses destroyed, 653 partly flooded, 500 houses transported by the tsunami and 41 death (Yoshimura, 2004)

**5. Conclusion**

The recordings in Agawa Pond allowed us to reconstruct the behaviour of the predecessors of the 2011 Tohoku-oki tsunami. The significance of these findings concerning the flow direction and the recording of backwash sequences must be restricted to the local topographic context of the pond itself and its surroundings.

The main conclusions are as follows:

- i) the topography controls the sedimentation processes that construct tsunami deposits, with the best preservation occurring in depressions (Wassmer et al., 2010). This is confirmed at Agawa Pond that recorded and preserved sedimentary signature of the four tsunami events that surge on Sendai Plain during the last 120 years while Goto et al., (2012) consider that there is only one possible historical record of large tsunamis over the last thousand years on the Sendai and adjacent plains (1611 Keicho tsunami).

- ii) thanks to the AMS technique, we show that within the tsunami sequences recorded, three sub-sequences have been identified for the Meiji tsunami and two for the Showa tsunami, indicating multiple energy pulses occurred during these events. A single pulse has been evidenced for Chilean tsunami. The Tohoku-oki tsunami deposit is much thinner relative to its predecessors and does not show sub-sequences or indicates multiple energy pulsations characterizing this event.

- iii) the double origin of the sands evidenced by Szczuciński et al. (2012) and Goto et al. (2012) for the Tohoku-oki tsunami is confirmed and this origin and the evolution of the relative proportions of seashore and dunes / terrigenous material evolved in the same way for the past tsunamis according to the different sequences recorded.

- iv) the foraminifera assemblage found in the deposits indicates an origin from Inner and Middle Sub-littoral Zone (ISZ and MSZ). The deposit of theses foraminifera is strongly related to the deposit of the fine sands. 65.1% of the total amount of foraminifera are deposited within sands of mean grain size ranging from 200 to 400 µm.

- v) The sediment supply being considered as constant for the successive events, the thickness of the deposits seems to be more related to the efficiency with which the pond plays a sediment trap role than to the relative energy of the event. Regarding the magnitude of the tsunami events, the classification from largest to smallest is as follows: (1) Tohoku-oki; (2) Meiji; (3) Chilean; and (4) Showa. In contrast, the Agawa Pond tsunami sequences may be classified according to decreasing deposit thickness as follows: Meiji (42 cm); Showa (27 cm); Chilean (17); and Tohoku-oki (14 cm). This may, in part, be explained by the sediment trapping efficiency of the pond decreasing with increasing sediment infill over time. The thickness of the deposits is, thus, affected by the antecedent conditions within the pond.

- Another aspect this study has revealed, is potentially the importance of the stock of material defining the importance of the tsunami deposit. Indeed, often tsunami deposits are associated with the magnitude of tsunamis, but for paleo-tsunami researchers don't take into account the evolution of the shore-line and the availability of material that can have an important impact on the amount of sediment that can be remobilized and consequently deposited. For Agawapond, a diachronic analysis of the evolution of the shoreline and the coastal plain shows that during the 20th century the environment has rapidly changed from a rice-field environment with sparse villages into a semi-urban mesh with denser road network and most especially a large international harbour, which disturbed the lagoon and the shoreline where a large amount of sediment could be collected by tsunami. Subsequently, the progressive reduction of tsunami deposit thickness with time is certainly to be attributed to the change in the depositional environment around Agawa-pond during the last decade (Fig. 8), certainly providing with a point of caution for future research on tsunami deposits and paleo-tsunami deposits.

As mentioned above in the text, the Meiji tsunami sequence is not complete in the core. A further investigation would be useful to find the base of this sequence and, may be, some older tsunami deposits recorded in Agawa Pond and some other ponds in the same area.